%% file: ICSE-Main.tex
\documentclass[conference]{IEEEtran}
\IEEEoverridecommandlockouts
\usepackage{cite}
\usepackage{amsmath,amssymb,amsfonts}
\usepackage{graphicx}
\usepackage{textcomp}
\usepackage{xcolor}
\def\BibTeX{{\rm B\kern-.05em{\sc i\kern-.025em b}\kern-.08em
    T\kern-.1667em\lower.7ex\hbox{E}\kern-.125emX}}

\input{definition}

\begin{document}

\title{AgentRaft: Detecting Data Over-Exposure in LLM Agents via Fine-grained Prompt Generation and Data Flow Inspection
\thanks{*Corresponding author: Jiangrong Wu.}
}

\author{\IEEEauthorblockN{Yixi Lin}
\IEEEauthorblockA{
\textit{Sun Yat-sen University}\\
Zhuhai, China \\
linyx98@mail2.sysu.edu.cn}
\and
\IEEEauthorblockN{Jiangrong Wu\textsuperscript{*}}
\IEEEauthorblockA{
\textit{The Hong Kong University of Science and Technology}\\
Hong Kong, China \\
jiangrongwu@cse.ust.hk}
\and
\IEEEauthorblockN{Xueqiang Wang}
\IEEEauthorblockA{
\textit{University of Central Florida}\\
Orlando, FL, USA \\
xueqiang.wang@ucf.edu}
\and[\hfill\mbox{}\par\mbox{}\hfill]
\IEEEauthorblockN{Xinyuan Zhang}
\IEEEauthorblockA{
\textit{National University of Singapore}\\
Singapore \\
zhangxinyuan@u.nus.edu}
\and
\IEEEauthorblockN{Yuhong Nan}
\IEEEauthorblockA{
\textit{Sun Yat-sen University}\\
Zhuhai, China \\
nanyh@mail.sysu.edu.cn}
}

\maketitle

\begin{abstract}

The rapid integration of Large Language Model (LLM) agents into autonomous task execution has introduced significant privacy concerns within cross-tool data flows. In this paper, we systematically investigate and define a novel risk in LLM agent termed \textbf{Data Over-Exposure (DOE)}, where an Agent inadvertently transmits sensitive data beyond the scope of user intent and functional necessity. We identify that DOE is primarily driven by the broad data paradigms in tool design and the coarse-grained data processing inherent in LLMs.

To this end, we present \textbf{\system{}}, the first automated framework for detecting DOE risks in LLM agents. The key challenge in detecting DOE lies in synthesizing targeted, risk-revealing user prompts that reliably trigger privacy-relevant cross-tool executions, and in determining whether the resulting sink payload exceeds the user's disclosure boundary. \system{} addresses this by combining program analysis with semantic reasoning: it builds a Cross-Tool Function Call Graph that guides both the synthesis of prompts triggering deep cross-tool executions and the runtime data-flow inspection that identifies privacy violations.

We evaluate \system{} on controlled agents derived from 6,675 real-world tools, where it achieves 92.48\% end-to-end precision and covers 98.68\% of DOE-affected chains with only 150 prompts, demonstrating accurate and efficient navigation of complex cross-tool execution spaces. We further evaluate \system{} on five existing real-world agents in the same tool environment and identify DOE in 32.48\% of the source-to-sink paths (759 out of 2,337). Our research highlights that DOE is a systemic risk across current real-world agents.

\end{abstract}


\input{main-body}

\bibliographystyle{IEEEtran}
\bibliography{ref}

\end{document}

%% file: definition.tex
\usepackage{cite}
\usepackage{amsmath,amssymb,amsfonts}
\usepackage{graphicx}
\usepackage{textcomp}
\usepackage{xcolor}
\usepackage{url}
\usepackage{xspace}

\usepackage{booktabs}
\usepackage{colortbl}
\usepackage{multirow}
\usepackage{makecell}
\usepackage{tabularx}
\usepackage{array}

\usepackage{algorithm}
\usepackage{algpseudocode}

\usepackage{fancyvrb}
\usepackage{mdframed}
\usepackage[most]{tcolorbox}
\tcbuselibrary{breakable}

\usepackage{enumitem}
\usepackage{pifont}

\usepackage[hidelinks]{hyperref}
\usepackage[capitalise,nameinlink]{cleveref}
\crefname{figure}{Fig.}{Figs.}
\Crefname{figure}{Fig.}{Figs.}

\newcommand{\para}[1]{\noindent\textbf{#1.~}}
\newcommand{\system}{\textit{AgentRaft}\xspace}

\newcommand{\Description}[1]{}

\newtcolorbox{graybox}{
    colback=gray!20,
    colframe=white,
    boxrule=0pt,
    width=\linewidth,
    left=1em,
    right=1em,
    top=0.5em,
    bottom=0.5em,
    breakable
}

%% file: main-body.tex
\section{Introduction}

The evolution of Large Language Models (LLMs) has catalyzed a paradigm shift in artificial intelligence, transitioning from passive conversationalists to \textit{action-oriented agents} capable of autonomous task execution~\cite{Code_Reviewer, Poster, Assistants_to_Agents}. By integrating with external tool ecosystems, these agents serve as a ``sophisticated brain,'' orchestrating closed-loop workflows that involve interpreting user intentions, planning sub-tasks, and coordinating data flow across diverse tools. This capability allows agents to handle complex requests, such as retrieving a file~\cite{file_agent}, extracting specific content, and emailing it to a colleague~\cite{email_agent}, by chaining multiple tool invocations (e.g., \texttt{read\_file} $\rightarrow$ \texttt{send\_email}). However, this architectural advancement introduces a complex, multi-stage data flow that inherently heightens the risk of unintended data exposure during autonomous execution.

In this paper, we refer to this privacy risk as \textbf{Data Over-Exposure (DOE)} in LLM Agents. The risk arises when an LLM agent transmits data outside both the user's intended sharing scope and the functional needs of the task during autonomous execution. Two characteristics of today's LLM agents increase the risk of DOE. First, to maximize flexibility, integrated tools often return a broad range of data regardless of whether it is necessary for the task. Second, LLMs lack contextual privacy awareness. Prior research~\cite{mireshghallah2023can,li20251} shows that LLMs frequently fail to judge which data should not be exposed in complex tasks, even when individual data points can be flagged as sensitive.

We use the following example to illustrate the problem of DOE. Consider Pokee~\cite{over-exposure-example}, a real-world LLM agent app in which a user asks the agent to ``extract the payment date from a transaction log and email it to an auditor.'' While the user only wants to share the date, the upstream tool (\texttt{read\_file}) returns a full data schema containing sensitive financial details. At the same time, the LLM fails to enforce a strict data boundary and inadvertently passes the whole record (e.g., credit card number and CVV) to the downstream tool (\texttt{send\_email}). As a result, a third party receives significantly more data than intended~\cite{gdpr, ccpa, pipl}.

While similar DOE issues have been extensively discussed in traditional systems such as mobile apps~\cite{tradition_software_leak_1, related_work_mobile_3} and IoT platforms~\cite{tradition_software_leak_5}, detecting such risks in LLM agents poses unique challenges to existing techniques. 
For example, prior research~\cite{related_work_program_analysis_3, tradition_software_leak_1} often analyzes prescribed lines of code of software with deterministic data processing and data flows.
However, in LLM agents, sensitive data usage is driven by \textit{dynamic, non-deterministic tool orchestration}, where the LLM determines which tools to invoke and how to process data at runtime.
As a result, techniques that rely on modeling data flows from static software code would likely fail, leaving dynamic analysis as the most viable approach to understand and monitor data flows.
Unfortunately, unlike traditional software, where test cases for triggering execution are often available or documented, creating a fixed set of test cases for LLM agents is not only time-consuming but also difficult, due to the unconstrained and probabilistic nature of tool executions.

\para{Our Work} In this paper, we propose the first automated, generic framework for detecting data over-exposure risks in LLM agents. In particular, given an Agent, we aim at generating a rich set of valid user prompts that can fully trigger various tools provided by the agents, and further confirm if there is any potential user unintended data exposure during the agent usage. 

To achieve this goal, three primary technical challenges need to be addressed. First, we need to comprehensively model the interaction landscape across various tools of Agent, as complex, multi-tool combinations can create hidden data-flow leakage channels. Second, we must synthesize high-validity user prompts that act as deterministic triggers to drive the agent through specific execution paths for valid DOE discovery. Finally, we need to accurately observe potential data exposure and distinguish between data that is functionally necessary for a tool and data that is genuinely over-exposed. 

To address these challenges, we adopt the concepts of program analysis, specifically call graph construction and data-flow analysis, to build \system{}. Our framework formalizes the agent's tool interactions into a structured execution space and operates through three synergistic stages. First, to capture the hidden data-flow of the toolset, \system{} constructs a Cross-Tool Function Call Graph (FCG) by modeling agent tools as inter-dependent function calls. This stage maps the underlying data-flow dependencies and transitive relationships across various tools and scenarios. Second, \system{} traverses the FCG to identify reachable execution call chains and synthesizes high-fidelity user prompts. These prompts act as deterministic triggers that drive the agent to execute specific, multi-step tool invocations. Finally, \system{} performs runtime data-flow tracking during execution. The framework can observe any data transmissions by capturing fine-grained data traces along the call chain. Moreover, to accurately distinguish between functional necessity and over-exposure, \system{} implements a multi-LLM voting mechanism grounded to assess whether observed data transmissions constitute over-exposure.

We evaluate \system{} on four controlled, sim-to-real Agents based on analyzing the usage of 6,675 top-popular tools from MCP.so. These agents covering mainstream usage scenarios (with popular tools), including Enterprise Collaboration, Software Development \& IT Operations, Social Platform Communication and Data Management \& Analysis (see Section \ref{sec:Evaluation} for more details). The evaluation measures two aspects: the end-to-end precision of \system{}, and its coverage of the ground-truth DOE call chains. We use a set of non-guided prompting as the baseline and compare it with \system{}.
Overall, \system{} achieves 92.48\% end-to-end precision in terms of detecting DOEs. More importantly, whereas non-guided baselines cover less than 20\% of the ground-truth DOE call chains after 300 prompts, \system{} reaches 98.68\% coverage with only 150 prompts and reduces the per-chain verification cost by 88.1\%.

In addition, our validation on existing real-world agents further shows that DOE is indeed a systems-level risk. More specifically, we run \system{} under a fixed backbone model and tool environment, and analyze 2,370 source-to-sink tool-call paths across five agents (e.g., Codex, Claude Code and Openclaw), covering the aforementioned four application scenarios. Among the 2,337 call-chain tests that successfully trigger their target paths, \system{} finds that 759 (32.48\%) of the test instances are indeed with DOE. Furthermore, 79.57\% of the fields transmitted in these DOE-affected executions fall outside the scope of both user intent and functional necessity. These findings expose a persistent failure of data minimization in current agents and underscore the need for systematic, fine-grained DOE auditing.

\system{} provides practical security and privacy benefits for developers, platforms, and end-users. Developers can use the framework to perform systematic privacy vetting to identify and remediate data over-exposure risks before release.

\para{Contribution} In summary, this paper makes the following contributions:

\begin{itemize}[leftmargin=*]
    \item We perform the first investigation regarding \textit{Data Over-Exposure} risk specifically within the cross-tool data flows of LLM Agents. 
    
    \item We develop \system{}, an automated framework that leverages program analysis to detect DOE risks in LLM Agents.
    \item We demonstrate that \system{} enables accurate and cost-effective DOE auditing on controlled agents derived from real-world tools, substantially outperforming non-guided baselines in reaching deep DOE-affected execution paths.

    \item We validate DOE across five existing real-world agents and reveal that DOE persists across agents and application scenarios, exposing a pervasive agent-level privacy risk in current LLM-agent systems.

\end{itemize}

\section{Background and Problem Statement}
\label{sec:background}

\subsection{LLM Agent Architecture} An LLM agent couples a language model with external tools, turning a conversational model into an action-oriented system. Given a task, it follows a ``Reasoning-Action-Observation'' loop: the LLM plans the next step, invokes suitable tool functions, and incorporates the returned observation to continue or terminate. Tools thus serve as the interface to external environments, exposing functions that retrieve user-controlled resources, process data, or transmit results to external recipients.

\begin{figure}[htbp]
    \centering
    \includegraphics[width=\linewidth]{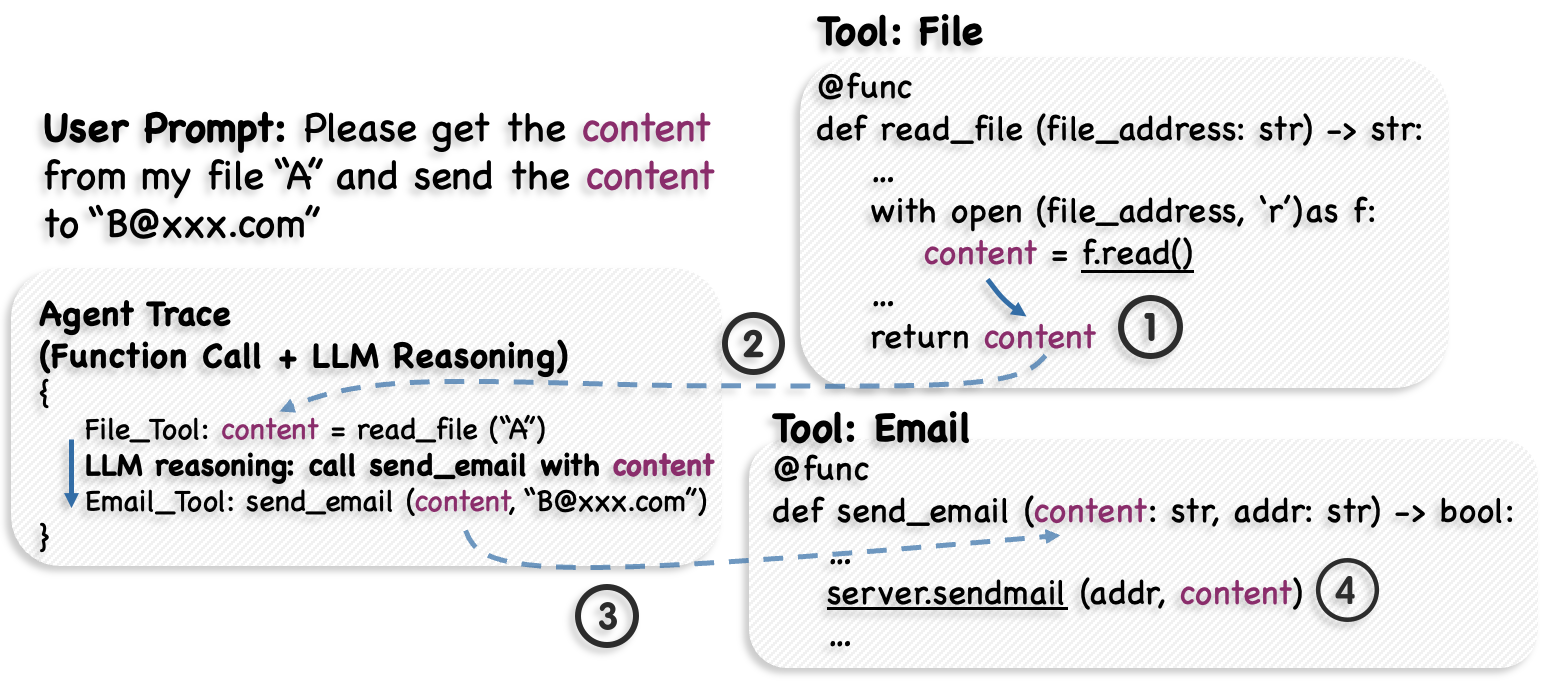}
    \caption{An example of cross-tool interaction with LLM reasoning.}
    \label{fig:data flow code example}
\end{figure}

\para{Cross-tool Interactions} The distinctive capability of an LLM agent lies in its ability to compose multiple tool functions into a coherent execution chain. As illustrated in \Cref{fig:data flow code example}, the LLM acts as a semantic mediator between heterogeneous tools: it may first invoke a retrieval function (e.g., \texttt{read\_file}) to obtain raw user data, extract task-relevant content from the returned observation, and then supply that content as an argument to a downstream function (e.g., \texttt{send\_email}). While this ``output-to-input'' dependency is fundamental for executing complex, multi-step tasks, it inherently creates a risk of data over-exposure, as private information retrieved at the source may be transmitted to an external sink beyond the user's original intent.

\subsection{Problem Statement}
\label{sec:data over exposure}

\begin{figure}[htbp]
    \centering
    \includegraphics[width=\linewidth]{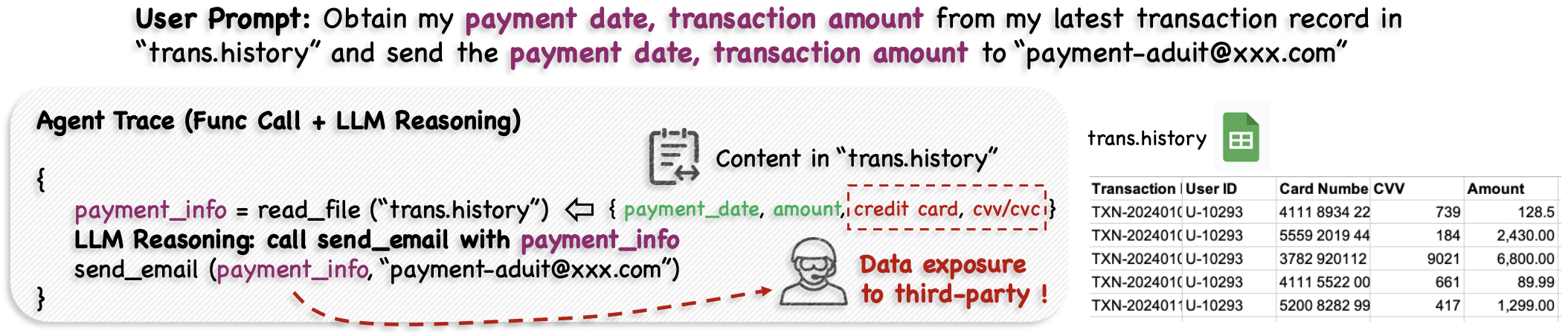}
    \caption{A real-world Data Over-Exposure scenario in LLM agent.}
    \label{fig:data exposure example}
\end{figure}

\para{Motivating Example} \Cref{fig:data exposure example} presents a real-world DOE case: the user asks the Agent to extract the payment date and transaction amount from a transaction record and send them to a third-party auditor. To complete the task, the Agent invokes a source function (e.g., \texttt{read\_file}) that returns the complete record, including the requested date and sensitive fields such as the credit card number and CVV. The LLM then constructs the input for a downstream sink function (e.g., \texttt{send\_email}) from this returned observation. Although only the payment date and transaction amount are intended for disclosure, the Agent forwards additional source-derived fields to the auditor, causing private data to cross the user's trust boundary. This example highlights the root cause of DOE in LLM agents: broad upstream tool outputs and LLM-mediated data transfer can obscure the boundary between task-relevant information and private data that should not be propagated.

\para{Definition and Scope of DOE}
We formally define Data Over-Exposure in LLM Agents by focusing on data transmissions that cross the user's trust boundary during normal task execution. We consider the user and the Agent components that process data exclusively on the user's behalf to reside within the same trust domain. A \textbf{\textit{third party}} refers to a distinct principal outside this domain that can observe the payload delivered by the Agent, such as an email recipient, collaborator, public audience, or external service outside the user's control. Accordingly, we define a \textbf{\textit{source}} function as a data-retrieval node that returns user-controlled data to the Agent (e.g., \texttt{read\_file}), and a \textbf{\textit{sink}} function as a terminal node that makes Agent-supplied data observable to a third party (e.g., \texttt{send\_email}). Data propagation among intermediate components within the same trust domain is therefore outside the scope of DOE.

DOE is assessed in the context of the current user request. A data transmission to a third-party sink does not inherently constitute a privacy violation, as the user may explicitly intend certain data to be shared. DOE arises when the sink payload contains source-derived data outside the scope of both the user's explicit transmission intent and the functional necessity of the requested task. Broad consent provisions, such as those in service providers' Terms of Service (ToS), are not used to determine task-specific disclosure boundaries, as their general descriptions of data use are insufficient to establish which disclosures are warranted by a particular user request.

Formally, given $D_{total}$ (all data retrieved at the source), $D_{trans}$ (source-derived data contained in the payload delivered to the sink), $D_{int}$ (data the user explicitly intends to transmit in the current task), and $D_{nec}$ (data not explicitly requested for transmission but semantically indispensable to fulfilling the user-requested task through the sink), the over-exposed data $D_{OE}$ is defined as:
\[
\scalebox{1.1}{\boldmath $D_{OE} = \left( D_{trans} \setminus (D_{nec} \cup D_{int}) \right) \cap D_{total}$}
\]

To this end, A DOE is identified when $D_{OE} \neq \emptyset$. Such data cross the user's trust boundary while falling outside both the user's explicit transmission intent and functional necessity.

\section{Overview of \system{}}

\para{Challenges and Solutions} 
To effectively detect DOE risks in a given LLM Agent, there are three challenges that need to be addressed:

\begin{itemize}[leftmargin=*]
    \item \textbf{C1: Highlighting valid tool usage combinations.} 
    DOE risks are often hidden within the vast space of potential tool combinations. Exhaustively testing all permutations is computationally prohibitive and frequently yields semantically invalid paths. The primary challenge lies in precisely modeling the semantic connectivity between heterogeneous tools to identify valid data dependencies across an expansive interaction space.
    
    \textit{\underline{Solution: Graph-based function call chain generation.}} We construct a cross-tool Function Call Graph (FCG) to formalize the Agent's potential tool interaction landscape. By employing a hybrid strategy of efficiency-driven static type-pruning and precision-oriented LLM validation, we provide a structural blueprint of reachable data-flow channels, ensuring hidden risks in diverse tool combinations are captured.

    \item \textbf{C2: Generating valid prompts along the execution path.} 
    Even with identified dependency tool call chains, the probabilistic nature of LLM decision-making makes guiding an Agent through a specific $source \rightarrow sink$ sequence difficult. Standard user prompts often fail to activate complex, multi-step tool chains required for risk validation.
    
    \textit{\underline{Solution: Call chain-driven prompt synthesis.}} We transform FCG paths into high-fidelity user prompts by instantiating nodes with specific assets and logical constraints. These prompts serve as deterministic triggers to enforce intended execution paths and validate complex call chains.

    \item \textbf{C3: Distinction between necessary data propagation and data over-exposure.} 
    Identifying privacy violations is non-trivial because the boundary of ``necessary data'' is dynamic and task-dependent. Transmitted data may exceed the user's explicit intent ($D_{int}$) yet remain strictly required for the tool to fulfill the specific task requested by the user ($D_{nec}$). Rigid auditing fails to capture this nuance, leading to high false-positive rates.
    
    \textit{\underline{Solution: Multi-model vote-based auditing.}} We implement a runtime module utilizing a committee of LLMs to analyze intercepted data ($D_{trans}$) alongside user intent and tool metadata. Here, the committee semantically delimits the $D_{nec}$ boundary (Data minimization principle), ensuring only data falling outside both $D_{int}$ and $D_{nec}$ is flagged as $D_{OE}$.
\end{itemize}

\begin{figure}[htbp]
    \centering
    \includegraphics[width=\linewidth]{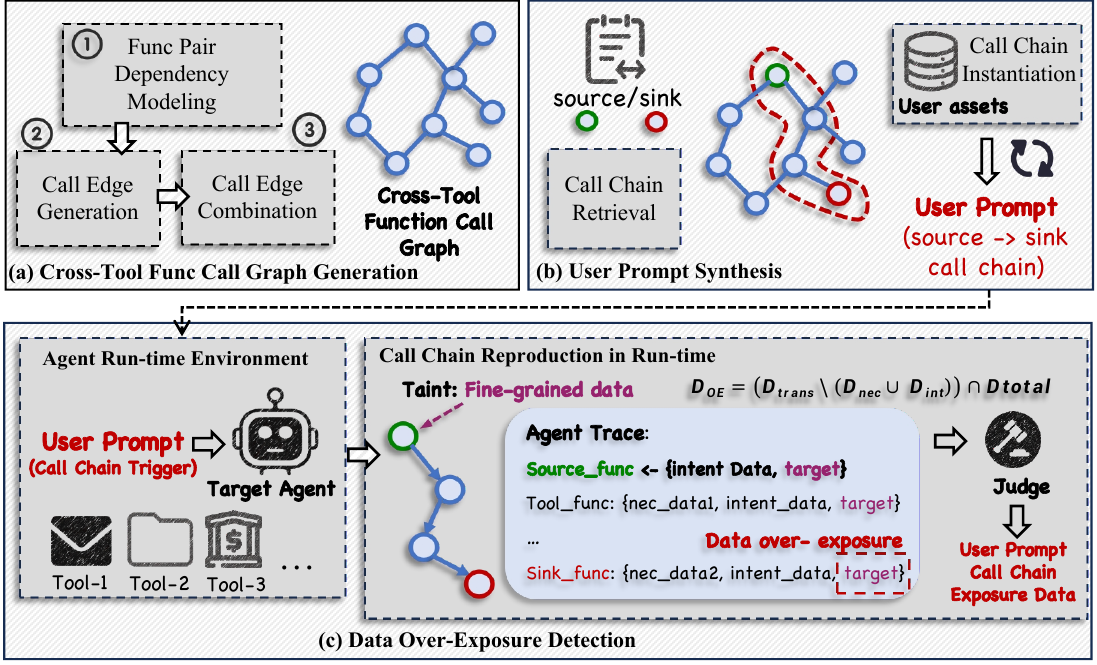}
    \caption{Workflow of \system{}.}
    \label{fig:workflow}
\end{figure}

\para{Overall Design}
As shown in \Cref{fig:workflow}, \system{} comprises three modules. \textit{Cross-Tool Function Call Graph Generation} models semantic data dependencies among Agent tools to identify valid source-to-sink paths. \textit{User Prompt Synthesis} transforms these paths into natural user prompts with explicit intent boundaries to trigger targeted multi-tool executions. \textit{Data Over-Exposure Detection} tracks runtime data flows and identifies source-derived fields that reach third-party sinks while falling outside both user intent and functional necessity. Together, these modules enable systematic, intent-aware DOE auditing of dynamic Agent workflows.

\section{Core Design of \system{}}

\subsection{Cross-Tool Function Call Graph Generation}
\label{subsec:agent data dependency construct}

\begin{figure}[htbp]
    \centering
    \includegraphics[width=\linewidth]{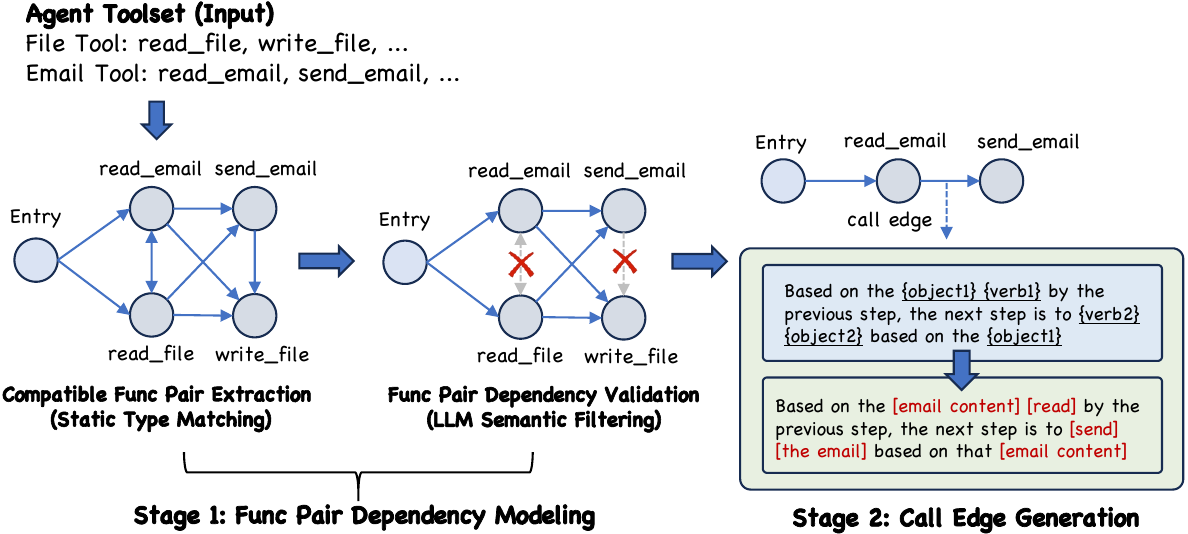}
    \caption{The process of Cross-Tool Function Call Graph generation.}
    \label{fig:cg generation}
\end{figure}

The primary objective of this module is to construct the data-flow landscape of the Agent system by identifying inter-tool and intra-tool function dependencies. \Cref{fig:cg generation} illustrates the whole process of this module.

\para{Cross-Tool Function Call Graph} Formally, we define $G = (N, E)$ as a directed graph representing the potential interaction landscape of an Agent's toolset. Unlike traditional call graph in software engineering that describe fixed, hard-coded execution paths, the cross-tool Function Call Graph (FCG) provides a systematic abstraction of potential semantic data dependencies across the entire toolset. While individual tools act as atomic operators for specific sub-tasks, the transitions between them are dynamically orchestrated by the LLM's reasoning. To this end, the FCG aims to capture all possible logical associations that the LLM might navigate during runtime. The FCG contains the following key components:
\begin{itemize}[leftmargin=*]
    
    \item \textbf{Entry Node}: A virtual node prepended to a call chain to represent the initial user-triggered action. Its outgoing edge encodes the instruction for invoking the first function.
    \item \textbf{Function Node}: Each function node $n \in N$ represents an individual function within the Agent's toolset (e.g., \texttt{read\_email}).
    \item \textbf{Call Edge}: An edge $e=(n_i,n_j)$ denotes a semantic data dependency in which the output of $n_i$ can supply data required by $n_j$. An edge from \textit{Entry} instead represents the initial user-triggered action for invoking the first function. Each edge is associated with an \textit{abstract action instruction} for subsequent prompt synthesis. For example, a chain from \textit{Entry} $\rightarrow$ \textit{read\_email} $\rightarrow$ \textit{send\_email} is connected by edges representing the logic of ``reading content of the email'' and ``transmitting that content via email.''
\end{itemize}

Based on this graph abstraction, \system{} constructs the FCG in two phases. \textit{Function Pair Dependency Modeling} identifies valid data dependencies through static analysis and LLM validation, while \textit{Call Edge Generation \& Combination} encodes these dependencies as instructed Call Edges and integrates them into the complete FCG.

\subsubsection{Function Pair Dependency Modeling} 
To accurately identify data dependencies between tool functions while controlling overhead, \system{} employs a hybrid strategy of \textit{Static Function Analysis and LLM-based Validation}. While a naive LLM-only approach could theoretically determine associations, it suffers from prohibitive token costs and latency when dealing with extensive toolsets.

\textbf{Stage 1: Compatible Pairs Extraction.} We first perform rapid static analysis on function signatures to identify compatible function pairs of the whole tool function set, and eliminate pairs that never transmit data. Specifically, \system{} checks whether the return type of one function and the input type of another satisfy the following relationships. (1) \textit{Type equivalence}: The input type of one function matches the return type of another; (2) \textit{Type subset}: The input/return type of one function is a subset of the return/input type of another; (3) \textit{Type conversion}: The return type of one function can be converted to the input type of another. If a function pair follows the above rules, \system{} flags it as compatible. For custom objects and containers, \system{} recursively scans nested member variables to ensure no potential data-flow link is overlooked.

\textbf{Stage 2: Dependency Validation.} Since static analysis only infers ``type-level'' possibility, it may produce false positives where the semantics are logically disjoint (e.g., read\_email and read\_file in \Cref{fig:cg generation}). Equipped with strong natural language understanding, the \system{} leverages LLM to judge whether the output of a preceding function is logically relevant to serve as the input of a subsequent function, based on functional descriptions and parameter names (e.g., ``extract email body'', ``send text message''). This step filters false-positive dependencies generated by static analysis, ensuring accurate dependency relationships. Ultimately, \system{} outputs all function pairs with data dependencies.

\subsubsection{Call Edge Generation \& Combination}
\label{para:call edge generation}
Once valid function pairs are identified, \system{} represents their dependencies as Call Edges. For each edge, \system{} generates an abstract action instruction to guide the corresponding function calls. To avoid semantic drift and ensure consistent path activation, we adopt a ``Semantic Extraction + Structured Template'' strategy. 
First, \system{} extracts the core \textit{Verb} (action) and \textit{Object} (target) from each function's natural language description. Then, \system{} fills these entities into a deterministic template to generate the abstract action instruction associated with each Call Edge. \textbf{\textit{Action Instruction Template}}: \textit{``Based on the \underline{\{object1\} \{verb1\}} by the previous step, the next step is to \underline{\{verb2\} \{object2\}} based on the \underline{\{object1\}}.''} 

For instance, the abstract action instruction associated with the call edge for \textit{read\_email} $\rightarrow$ \textit{send\_email} is structured as: \textit{``Based on the \underline{[email content]} \underline{[read]} by the previous step, the next step is to \underline{[send] [the email]} based on that \underline{[content]}.''} This structured generation ensures that the resulting user prompts are precise enough to drive the Agent through the intended data-flow path without ambiguity. Finally, \system{} integrates all nodes and validated Call Edges into a comprehensive Function Call Graph for downstream prompt generation and data over-exposure detection.

\subsection{User Prompt Synthesis}
\label{subsec:call chain user prompt synthesis}

To effectively detect data over-exposure, \system{} must ensure that the Agent not only has the capability to interact with tools but is also deterministically driven to traverse specific ``source-to-sink'' paths. This module takes the FCG and the source
and sink functions of interest as inputs, retrieves the
corresponding source-to-sink paths, and transforms them into
executable, high-fidelity user prompts. The synthesis process includes the following two stages.

\begin{figure}[htbp]
    \centering
    \includegraphics[width=\linewidth]{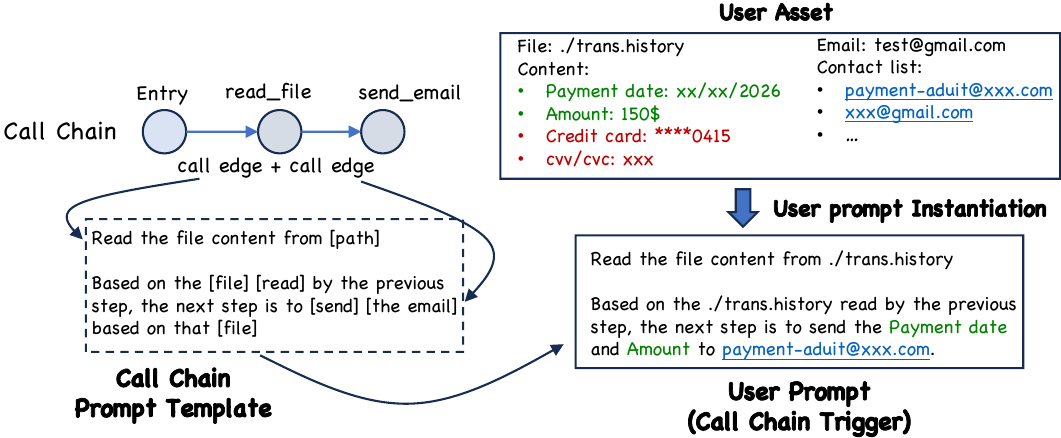}
    \caption{The process of User Prompt Synthesis.}
    \label{fig:user prompt synthesis}
\end{figure}

\subsubsection{Call Chain Retrieval}

The primary goal of this stage is to exhaustively extract all reachable execution call chains from a designated \textit{source} node to a \textit{sink} node, and further generate the prompt template of the call chain. Given the FCG constructed in the previous module, we employ a \textit{Breadth-First Search (BFS)} algorithm to traverse the graph and identify all acyclic paths. 
The retrieval process validates the existence of the source-sink pair and maintains a queue of explored functional chains. To ensure testing efficiency and prevent infinite loops, we explicitly prune cyclic trajectories. After that, \system{} obtains a set of call chains from source to sink, each chain representing a potential data-flow channel that requires privacy validation.

Furthermore, as a call chain is essentially constituted by a sequence of validated call edges, \system{} synthesizes a dedicated call chain prompt template by concatenating the individual instructions of these constituent edges (\Cref{fig:user prompt synthesis}). By merging the standardized semantic templates of each edge from the call chain (\Cref{para:call edge generation}), \system{} constructs a logically coherent workflow instruction template. This resulting Call Chain Prompt Template serves as the foundational structure for subsequent user prompt instantiation. It ensures that the final trigger accurately encapsulates the multi-step execution logic required to drive the Agent through the entire data-flow call chain from source to sink. Each retrieved call chain therefore serves as an independent end-to-end test specification, from which \system{} synthesizes prompts with chain-specific user intent ($D_{int}$) and concrete assets. Structurally overlapping call chains may thus represent distinct task contexts and are evaluated separately.

\subsubsection{User Prompt Instantiation along the Call Chain} 

While a retrieved call chain prompt template provides a structural workflow description, it remains an abstract template ``skeleton.'' This stage involves transforming the prompt template into a \textit{concrete natural language user prompt} that an Agent can execute. 

As shown in \Cref{fig:user prompt synthesis}, \system{} first performs fine-grained labeling on the user's accessible assets, such as local files or contact lists. For each test case, a subset of data fields is designated as user intent data ($D_{int}$), representing the specific information to be explicitly requested for transmission (e.g., \textit{Payment date} and \textit{Amount}). The remaining fields from the same source asset are retained as over-exposure candidates (e.g., \textit{Credit card number} and \textit{CVV/CVC}). Since this partition is established before prompt generation, $D_{int}$ provides an explicit and deterministic intent boundary for the test case.

\system{} then employs an LLM to resolve abstract placeholders in the call-chain template. Using the function metadata, labeled $D_{int}$, and corresponding user assets, the LLM produces concrete prompts that guide the Agent along the intended source-to-sink call chain. During instantiation, \system{} further requires each generated request to be grounded in the selected asset and expressed as a purposeful user task, so that the intended data transmission is triggered through a natural user need rather than a mechanical tool-invocation instruction. For robust testing, each call chain is tested with a configurable number of prompts (five in our evaluation) and reported as DOE-affected if any prompt execution yields a non-empty $D_{OE}$.

\subsection{Data Over-Exposure Detection}

The final module of \system{} performs a quantitative assessment of privacy integrity by monitoring the Agent’s runtime behavior and verifying data boundaries. As illustrated in \Cref{fig:workflow}(c), this module is designed to detect whether unauthorized sensitive data leaks through the function call chain via the data flow tracking mechanism and automated judging pipeline. Finally, \system{} reports all call chains affected by DOE risk.

\subsubsection{Taint Data Tracking} 
In this phase, \system{} executes the synthesized prompts within a controlled Agent Runtime Environment to reproduce the execution path. The core of our tracking lies in the fine-grained labeling of data retrieved at the \textit{source} function. Specifically, once the \textit{source} function returns the full dataset $D_{\text{total}}$, \system{} identifies data elements that fall beyond the specified user intent ($D_{\text{int}}$) and assigns them a \textit{Taint Label} (marked as \textit{target} in our trace). It marks a candidate for subsequent necessity assessment, rather than a confirmed DOE.

To provide a deterministic view of data propagation, \system{} captures payloads at three critical observation points through the Agent Trace (i.e., Agent execution log): \textbf{\textit{(i) the Source\_function}}, to ensure tainted data and $D_{\text{int}}$ are correctly isolated; \textbf{\textit{(ii) the Tool\_function}}, to track intermediate processing; and \textbf{\textit{(iii) the Sink\_function}}, to intercept the final payload delivered to a third party across the user's trust boundary. Since source-derived fields may be paraphrased, reformatted, or combined during execution, \system{} further leverages an LLM to perform semantic correlation across the recorded payloads, identifying transformed representations of source-derived fields and marking them accordingly. This process enables fine-grained tracking of data propagation throughout the functional chain.

\subsubsection{Over-Exposure Judging}
\label{subsubsec:over exposure judge}

To determine whether transmitted data $D_{\text{trans}}$ constitutes a violation, \system{} performs field-level necessity analysis using a multi-LLM voting committee. The committee is guided by data-minimization criteria derived from three representative privacy regulations, GDPR~\cite{gdpr}, CCPA~\cite{ccpa}, and PIPL~\cite{pipl}. Data is considered over-exposed only if it falls outside the union of user-intended data ($D_{int}$) and functionally necessary data ($D_{nec}$).

Since $D_{int}$ is explicitly established during prompt synthesis, the committee focuses on transmitted candidate fields outside this intent boundary. For each candidate field, the LLM judges are provided with the synthesized user prompt, the sink function's metadata, and the intercepted sink payload. To assess functional necessity, each judge determines whether omitting the field from the sink payload would prevent the fulfillment of the original user request. A field is considered functionally necessary only if its transmission is indispensable to task completion, rather than merely relevant to the task or available in the source data. The majority vote determines whether each candidate field belongs to $D_{nec}$. A transmitted source-derived field is identified as DOE only if it falls outside both $D_{int}$ and $D_{nec}$. The reliability of this voting mechanism is evaluated in \Cref{subsec:rq2}.

\input{Material/Tables/Detailed-Characteristics-of-Major-LLM-Agent-Application-Scenarios}

\section{Evaluation}
\label{sec:Evaluation}

\subsection{Evaluation Setup}
\label{subsec:evaluation setup}
\para{Dataset} 
To systematically evaluate Data Over-exposure (DOE) risks in the current LLM Agent ecosystem, we construct a testing environment based on real-world tools. More specifically, we crawled \textbf{6,675} tools from MCP.so~\cite{mcpso}, one of the largest markets for the Agent tool. Our analysis revealed that the ecosystem is dominated by four primary scenarios: \textbf{Data Management \& Analysis} (58.5\%), \textbf{Software Dev \& IT Ops} (56.6\%), \textbf{Enterprise Collaboration} (39.8\%), and \textbf{Social Platform Communication} (21.7\%). The detailed distribution is available in our code repository.

To better simulate the usage of real-world tools, we construct four representative Agents (\Cref{table:agent_scenarios_final}) that cover four dominant agent application scenarios. For each Agent, we use an LLM to curate a scenario-specific MCP toolset that covers the common functional requirements of the corresponding domain. These four Agents serve as the controlled evaluation subjects for the component analysis, ablation study, and overhead measurement (RQ1--RQ4), while RQ5 further evaluates \system{} on five existing real-world agents.

\para{Manual Annotation}To establish reliable ground truth for our evaluation, we adopt a double-annotation protocol for three tasks: FCG construction, DOE-chain verification, and field-level DOE identification. Two researchers independently annotate the same samples for each task following predefined criteria. We measure inter-annotator agreement using Cohen's Kappa \cite{cohen1960coefficient}, obtaining values of 1.0, 0.98, and 0.89 for the three tasks, respectively. Disagreements are first resolved through discussion, with any remaining cases adjudicated by a third researcher.

\para{Environment Setting}We instantiate the four controlled Agents and evaluate \system{} on AgentDojo~\cite{agentdojo}. The extensible architecture of AgentDojo allows us to configure scenario-specific toolsets and collect consistent tool-level execution traces, including the inputs and outputs of source, intermediate, and sink invocations. The standardized execution environment also enables reproducible ground-truth construction and controlled comparisons across individual components and ablated variants. We implement the representative tool functionalities according to the specifications collected from MCP.so.

For internal modules, we adopted a modular model strategy: DeepSeek-V3.2~\cite{deepseekv3_2} for \textit{Cross-Tool Function Call Graph Generation}, Qwen3-Plus~\cite{qwen3_plus} for \textit{User Prompt Synthesis}, and a voting committee (GPT-4.1~\cite{gpt4_1}, Qwen3-Plus~\cite{qwen3_plus}, DeepSeek-V3.2~\cite{deepseekv3_2}) for DOE detection. The details of model selection is in \Cref{subsec:rq2}. The tested agents are consistently powered by DeepSeek-V4-Flash~\cite{deepseek_v4_flash}. Note that the choice of specific LLM only represents different Agent implementation configurations and has nothing to do with the detection mechanism or effectiveness of \system{}. 

Regarding user assets, we leverage DeepSeek-V3.2 to automatically generate specific resources based on the environment schema and tools of each Agent. These assets are strategically populated with a controlled mixture of simulated Personally Identifiable Information (PII) and business data to provide a rigorous foundation for evaluating data over-exposure risks across the selected scenarios.

\para{Research Questions}
We evaluate \system{} by answering the following five research questions:
\begin{itemize}[leftmargin=*]
    \item \textbf{RQ1: Overall Effectiveness.} How effectively can \system{} identify Data Over-Exposure (DOE) risks?
    \item \textbf{RQ2: Effectiveness of Individual Components.} What are the specific contributions and performance of \system{}'s core modules?
    \item \textbf{RQ3: Ablation Study.} How does each component contribute to the overall performance of \system{}?
    \item \textbf{RQ4: Performance Overhead.} What are the computational and time costs of deploying \system{} for auditing?
    \item \textbf{RQ5: Real-world Validation.} How prevalent is DOE in real-world LLM agents?
\end{itemize}

\subsection{RQ1: Overall Effectiveness}

\para{Evaluation Setup}
We evaluate the end-to-end precision of \system{} at the call-chain level across the four controlled Agents. We test each call chain with five synthesized user prompts to improve the robustness of the evaluation. A call chain is reported as DOE-affected if at least one execution produces a non-empty $D_{OE}$. Following the formal definition of DOE in \Cref{sec:data over exposure}, we manually verify each reported chain by examining its synthesized user prompt, runtime execution trace, and sink payload. Specifically, annotators identify source-derived fields transmitted to a third-party sink and assess whether they fall outside $D_{\mathrm{int}} \cup D_{\mathrm{nec}}$. A reported chain is counted as a true positive if at least one flagged execution is manually confirmed to contain DOE; otherwise, it is counted as a false positive.

\input{Material/Tables/overall-result}

\para{End-to-End Effectiveness}
The end-to-end results demonstrate the reliability of the DOE findings reported by \system{}. As shown in \Cref{table:overall-results}, \system{} reports 319 DOE-affected call chains among the 474 tested chains, of which 295 are manually confirmed, yielding an overall precision of 92.48\%. Precision exceeds 90\% in all four scenarios and reaches 93.83\% in the \textit{Enterprise} scenario. These results indicate that FCG-guided path discovery, intent-constrained prompt synthesis, and multi-LLM necessity analysis jointly enable reliable DOE detection across diverse Agent application scenarios.

We do not report the end-to-end recall of \system{} due to the lack of an exhaustive ground truth. More specifically, establishing an exhaustive ground truth of DOE-affected call chains is impractical due to the open-ended space of user prompts and the nondeterministic nature of agent execution. To complement the end-to-end precision evaluation, we further assess the effectiveness of \system{}'s three core stages in RQ2.

\subsection{RQ2: Effectiveness of Individual Components}
\label{subsec:rq2}

\input{Material/Tables/CG-Construction}

\para{Robustness of Constructed FCG}
To assess the robustness of the FCG as the structural foundation of \system{}, we compare the generated graphs with manually established ground-truth graphs containing 233 nodes and 720 edges. Following the annotation protocol in Section \ref{subsec:evaluation setup}, annotators examine function inputs and outputs, parameter descriptions, and functional semantics to determine whether the output of one function can supply data required by another. A directed edge is considered valid when such a data dependency exists. As shown in \Cref{table:call-graph-results}, \system{} achieves an overall Precision of 93.96\%, Recall of 90.69\%, and F1-score of 92.30\%, demonstrating its ability to accurately capture tool-level data dependencies. In particular, the high recall indicates that the generated FCGs cover most valid interaction paths, thereby narrowing the search space for targeted prompt synthesis while reducing the risk of missing deep DOE paths. Our error analysis identifies 42 FPs and 67 FNs. FPs mainly arise when semantic similarities in tool descriptions, such as those of logging and data-retrieval utilities, lead to non-existent dependencies. In contrast, FNs typically result from implicit dependencies or non-standard data formats that are not clearly expressed in API signatures. Nevertheless, the overall results confirm that the generated FCGs provide a reliable blueprint for subsequent prompt synthesis.

\input{Material/Tables/User-Prompt-Generation}

\para{Quality of Synthesized User Prompts}
To evaluate the fidelity of prompt synthesis, we measure whether the generated prompts can successfully guide the Agent through their targeted tool call chains. As summarized in \Cref{table:prompt-generation-high-perf}, 2,207 of 2,370 prompts successfully trigger the intended execution paths, yielding an aggregate coverage of \textbf{93.12\%}. The coverage remains consistently high across all scenarios and reaches 98.61\% in the \textit{Data Management} scenario, showing that most synthesized prompts are executable and suitable for subsequent DOE inspection. The remaining 6.88\% fail primarily because of a semantic gap between the auditing intent and the Agent's runtime reasoning. Specifically, LLM reasoning deviations, including instruction drift~\cite{LLM_instruction_follow}, may cause the Agent to misinterpret the request or invoke an alternative tool. In addition, unmet path constraints or state dependencies, such as missing intermediate identifiers, may lead to premature termination or deviation from the targeted path. Despite these failures, the overall trigger coverage remains sufficient to support systematic DOE auditing across complex tool-call paths.

\para{DOE Identification} To evaluate the precision of our automated auditing, we assess the \textit{Over-Exposure Judge} (\Cref{subsubsec:over exposure judge}) module at data-field level. For each scenario, we randomly sample 50 manually verified DOE-triggering user prompts and the same number of manually verified non-DOE user prompts. Annotators examine the user prompts, runtime traces, and sink payloads to label transmitted source-derived fields according to the DOE criteria defined in Section \ref{subsubsec:over exposure judge}. These manually established labels serve as ground truth for evaluating the Over-Exposure Judge. As summarized in \Cref{table:doe-judge-sampled}, \system{} achieves an aggregate Precision of 97.48\%, a Recall of 98.25\%, and an F1-score of 97.86\%. The metrics remain consistently high across all evaluated domains, ranging from 97.28\% to 98.26\% F1-score, demonstrating that our framework can successfully distinguish between functionally necessary data ($D_{\text{nec}}$) and over-exposed fields with minimal false alarms.

\input{Material/Tables/DOE-Identification}

A comparative analysis (\Cref{table:doe-judge-comparison}) shows that our multi-LLM voting committee substantially outperforms individual models. Single-model baselines (GPT-4.1, DeepSeek-V3.2, Qwen3-Plus) achieve F1-scores of 88.89–90.68\%, whereas our consensus mechanism reaches 97.92\%, outperforming the best single-model judge by 7.24 percentage points. False positives drop from up to 22 to 4, demonstrating the committee's ability to reduce individual model bias and hallucinations. Remaining FPs and FNs (4 and 3) mainly arise from nuanced semantic edge cases where ``functional necessity'' remains inherently ambiguous under privacy regulations.

\input{Material/Tables/DOE-Identification-with-different-LLM}

\para{Impact of LLM Models} To justify our modular model selection, we evaluate different LLM backbones for the two modules of \system{} (\textit{FCG Generation} and \textit{User Prompt Synthesis}). As shown in \Cref{table:model_selection}, candidate models exhibit distinct advantages based on the specific cognitive requirements of each task. For \textit{FCG Construction}, which requires precise mapping of tool dependencies, DeepSeek-V3.2 achieves a peak F1 score of 96.45\% and identifies all 76 vulnerable call chains. Conversely, Qwen3-Plus performs best in \textit{User Prompt Synthesis}, reaching 92.04\% trigger coverage and generating 287 valid prompts that successfully activate DOE.

\input{Material/Tables/Impact-of-different-LLM-CG-and-UP}

This performance divergence further underscores the necessity of both the FCG and User Prompt Synthesis modules within \system{}. As DOE detection is a sequential pipeline, its effectiveness is bounded by FCG fidelity and prompt coverage. Specifically, low FCG accuracy would lead to an incomplete identification of potential attack paths, causing the system to overlook deep-seated risks (False Negatives). Similarly, insufficient prompt coverage may fail to activate correctly identified risks. Therefore, the modular employment of specialized models is essential to ensure that neither stage becomes a bottleneck, thereby maximizing the end-to-end detection rate of complex privacy risks in LLM Agents.

\begin{figure}[htbp]
    \centering
    \includegraphics[width=0.7\linewidth]{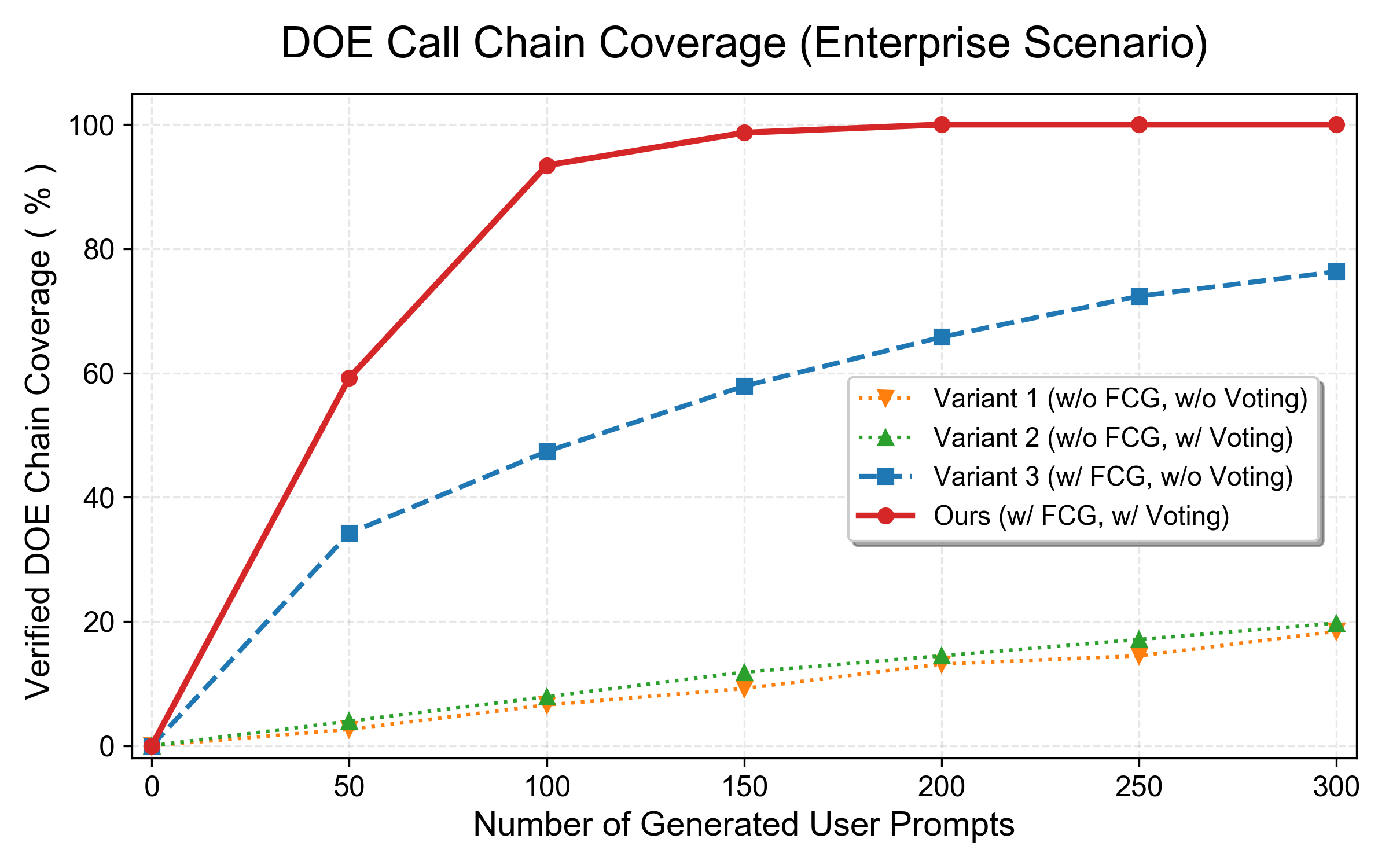}
    \caption{Ablation of FCG-guided prompt synthesis and multi-LLM voting.}
    \label{fig:baseline_compare}
\end{figure}

\subsection{RQ3: Ablation Study}

\para{Evaluation Setup} We conduct the ablation study using the 76 manually confirmed DOE-affected call chains from the \textit{Enterprise Collaboration} scenario in RQ1. These chains serve as a fixed positive benchmark for RQ3 and RQ4. We evaluate all variants against this benchmark under the same prompt budget, measuring coverage as the proportion of known DOE-affected chains identified by each variant. This setup allows us to assess their effectiveness in discovering DOE risks within a limited testing budget. We use a single scenario primarily due to cost constraints. The results are applicable to an arbitrary LLM agent or scenario. We compare three variants to examine how prompt generation and judging affect the results. \textbf{Variant 1} uses an LLM to randomly generate user prompts and employs GPT-4.1 as the sole judge for DOE identification. \textbf{Variant 2} adopts the same prompt generation method as Variant 1 but employs a multi-LLM voting judge. \textbf{Variant 3} uses FCG-guided prompt synthesis and retains GPT-4.1 as the sole judge. \Cref{fig:baseline_compare} shows the comparison results.

\para{User Prompt Generation} The results highlight the importance of the Function Call Graph (FCG) in navigating the Agent's tool-use logic. Without FCG, the two random-prompt configurations (\textit{Random + Single Judge} and \textit{Random + Voting Judge}) fail to activate deep-seated logical call chains, capturing only 18--20\% of risks even after 300 prompts. In contrast, \system{} captures 59.21\% of ground-truth DOE instances within the first 50 prompts. This underscores that FCG-guided synthesis effectively optimizes the search space, allowing for the rapid generation of high-quality test user prompts that target complex multi-tool sequences.

\para{Over-Exposure Judge} The necessity of consensus-based identification is evidenced by the gap between \system{} and the \textit{FCG-guided + Single Judge} configuration. Although this configuration uses FCG-guided synthesis, its discovery rate plateaus at 76.32\% due to single-model errors and hallucinations. \system{} overcomes this bottleneck via multi-LLM voting, achieving near-total coverage ($\sim$99\%) with 150 prompts. These results show that while FCG enables call chain activation, voting reduces identification noise and ensures high-precision DOE verification across diverse Agents.

\input{Material/Tables/overhead}

\subsection{RQ4: Performance Overhead}

We evaluate \system{}’s efficiency across temporal and economic dimensions using the same benchmark as in RQ3. Results are detailed in \Cref{table:cost-efficiency-final}.

As can be seen from the table, \system{} generates 150 prompts in 128.93s and completes detection in 1,635.23s, achieving 98.68\% coverage. Although multi-LLM voting introduces slight latency over the \textit{FCG-guided + Single Judge} variant (1,497.06s), it achieves substantially higher coverage, while non-FCG variants remain below 12\% within similar time. Auditing consumes 1.15M tokens, or 0.015M per verified chain, reducing per-chain cost by \textbf{88.1\%} over the \textit{Random + Voting Judge} baseline; thus, the reduced prompt volume offsets the voting overhead. The projected full-coverage cost is \textbf{1.2M} tokens, compared with \textbf{8.2M} tokens for the single-judge variant due to its false negatives, whereas random-prompt variants fail to converge within the evaluated budget.

\subsection{RQ5: Validating DOE in Real-World Agents}

\input{Material/Tables/realworld-results}

\para{Experimental Setup}
We further evaluate whether DOE appears in existing real-world agents rather than only in our self-crafted agents. Specifically, we instantiate five agents, Claude Code~\cite{claude_code}, Codex~\cite{codex}, Gemini CLI~\cite{gemini}, OpenClaw~\cite{openclaw}, and Hermes~\cite{hermes_agent}, in the same controlled tool environment. Each agent retains its own scaffold, system prompt, and tool-use policy while sharing the same backbone model (DeepSeek-V4-Flash). This setting controls for model-level differences and isolates the effect of agent-level data handling. For each scenario, we evaluate all five agents using the same set of source-to-sink call chains, with five synthesized user prompts per chain. Across all scenarios, this produces 2,370 call-chain tests and 11,850 prompt executions. To ensure a consistent comparison across agents, we calculate DOE rates at both the call-chain and prompt levels based on executions that successfully trigger the target source-to-sink call chain. \Cref{table:realworld-results} reports DOE at the call-chain, user-prompt, and data-field levels.

\para{DOE Prevalence}
As shown in \Cref{table:realworld-results}, DOE appears across all five agents and all scenarios. At the call-chain level, DOE occurs in 759 of the 2,337 call-chain tests with at least one successful target-chain execution (32.48\%). At the user-prompt level, 2,896 of the 11,009 executions that successfully trigger their target chains (26.31\%) exhibit DOE. These results suggest that DOE is not specific to a single agent or scenario. At the data-field level, among fields transmitted in DOE-triggering executions, 12,219 out of 15,356 fields (79.57\%) are over-exposed. Manual inspection of representative DOE traces further shows that the exposed fields include \textit{credential-like secrets}, \textit{authentication data}, and \textit{bulk customer PII}. These cases highlight the severity of DOE: ordinary source-to-sink executions can move high-impact private data across the user's trust boundary without explicit intent or functional necessity.

\para{Impact of Agent Design}
With the backbone model and tool environment held fixed, differences in DOE rates mainly reflect agent-level behavior. The variation is substantial. In the \textit{Social} scenario, prompt-level DOE ranges from 27.80\% for Claude Code to 65.12\% for OpenClaw under the same model and tools. Across all scenarios, Claude Code has the lowest prompt-level DOE rate (11.67\%), followed by Hermes (23.43\%), OpenClaw (30.62\%), Gemini (32.18\%), and Codex (33.60\%). This gap suggests that agent-level design choices, such as how much retrieved context is preserved, summarized, or forwarded to downstream tools, materially affect DOE risk. However, no agent eliminates DOE: even Claude Code triggers DOE in 27.80\% of prompts in the \textit{Social} scenario.

\section{Discussion}

\para{Application Scenarios of \system{}}
While presented as a detection framework, \system{}’s Cross-Tool Function Call Graph (FCG), test case generation, and data flow analysis provide a deterministic foundation for broader security tasks in the agent ecosystem. The FCG maps an agent's ``safe operating space'', offering a structural blueprint beyond auditing. For instance, the FCG can be employed for runtime anomaly detection, where any deviation in the execution trace from the statically verified edges could signify prompt injection or logic hijacking. Furthermore, it can assist pre-deployment policy enforcement by helping developers check whether validated paths exist between sensitive sources and third-party sinks before an agent is released. In this light, \system{} not only uncovers privacy risks but provides the primitives to make LLM agents auditable and trustworthy.

\para{Limitations} While \system{} provides a systematic approach to detecting DOE risks, several limitations remain. First, the overall effectiveness of our framework is intrinsically linked to the underlying LLM's reasoning performance, and occasional hallucinations or reasoning failures may affect the precision and recall of risk detection. Second, our FCG construction currently focuses on static and semi-dynamic dependency mapping; in environments where tools are dynamically added or modified at runtime, the graph may require frequent re-validation to stay accurate. Overall, these limitations stem from the current capabilities of LLMs and the dynamism of agent ecosystems rather than from the design of \system{} itself, and they point to clear directions for future work, such as adopting stronger reasoning backbones and extending the FCG to fully dynamic tool environments.

\section{Related Work}

\para{Privacy Leakage in LLM-based Systems}
The integration of Large Language Models (LLMs) into autonomous agents has expanded the attack surface for privacy violations~\cite{related_work_LLM_1, related_work_LLM_2, related_work_LLM_3}. Early research focused on model-centric risks, such as training data extraction~\cite{related_work_LLM_4, related_work_LLM_5} and membership inference attacks~\cite{related_work_LLM_6, related_work_LLM_7}. As agents began utilizing external tools, research shifted toward execution-time vulnerabilities~\cite{related_work_LLM_8, related_work_LLM_9, related_work_LLM_4, related_work_LLM_11, related_work_LLM_12}, including prompt injection~\cite{related_work_LLM_9, related_work_LLM_4, mail2027} and malicious plugin exploitation~\cite{related_work_LLM_11, related_work_LLM_12, related_work_LLM_13}, self-propagating attacks \cite{agentworm2027}, and over-privileged Agent Skills \cite{skillscope2026}. However, most existing works focus on adversarial scenarios where an attacker intentionally triggers a leak. Our work diverges by focusing on architectural risks where legitimate tool-calling sequences inadvertently cause data over-exposure during normal operations.

\para{Data Over-Exposure (DOE) in Traditional Software}
DOE is a well-recognized challenge in software engineering, particularly within Web APIs~\cite{related_work_web_1, related_work_web_2, related_work_web_3}, mobile application backends~\cite{related_work_mobile_1, related_work_mobile_2, related_work_mobile_3}, and IoT systems~\cite{related_work_iot_1, related_work_iot_2, related_work_iot_3}. In traditional systems, DOE occurs when a server returns a full database object to the client, even if only a subset of fields is required for the user interface. To mitigate this issue, prior work has proposed fine-grained data fetching mechanisms such as GraphQL~\cite{graphql}, as well as program analysis techniques~\cite{related_work_program_analysis_1, related_work_program_analysis_2, related_work_program_analysis_3, related_work_program_analysis_4} to detect over-sharing in REST APIs and privacy-sensitive applications.
Beyond API design, privacy compliance and automated auditing have also relied on static and dynamic analysis~\cite{related_work_privacy_compliance_1, related_work_privacy_compliance_2, related_work_privacy_compliance_3, related_work_privacy_compliance_4}, often leveraging formal methods and taint tracking~\cite{related_work_privacy_compliance_5, related_work_privacy_compliance_6, related_work_program_analysis_1, related_work_program_analysis_2, related_work_program_analysis_3, related_work_program_analysis_4} to monitor sensitive data flows in Web and mobile environments. More recent efforts further incorporate natural language processing to bridge the gap between regulations like GDPR and CCPA and system behaviors~\cite{related_work_privacy_compliance_7, related_work_privacy_compliance_8, related_work_privacy_compliance_9}.

\section{Conclusion}
This paper presents \system{}, which combines structural call graph modeling, user prompt synthesis, and multi-model consensus auditing to detect Data Over-Exposure (DOE) in LLM agents. \system{} achieves 92.48\% end-to-end precision and reduces per-chain verification cost by 88.1\% over non-guided baselines. Across five real-world agents, DOE occurs in 759 of 2,337 successfully triggered source-to-sink call-chain tests, and 79.57\% of fields transmitted in DOE-triggering executions are over-exposed.

%% file: Material/Tables/Detailed-Characteristics-of-Major-LLM-Agent-Application-Scenarios.tex
\begin{table*}[htbp]
\setlength{\belowcaptionskip}{0pt}
\caption{The four controlled Agents used in our evaluation, with their target scenarios, toolsets, and numbers of functions, sources, and sinks.}
\label{table:agent_scenarios_final}
\centering
\small
\setlength{\tabcolsep}{3.5pt}
\renewcommand{\arraystretch}{1.08}
\begin{tabularx}{\textwidth}{>{\raggedright\arraybackslash}p{0.28\textwidth}|>{\raggedright\arraybackslash}X|>{\raggedright\arraybackslash}p{0.25\textwidth}|c|c|c}
\Xhline{1.1pt}
\textbf{Agent name (Usage scenario)} 
& \textbf{Introduction} 
& \textbf{Toolset} 
& \textbf{\# Func} 
& \textbf{\# Source\textsuperscript{1}} 
& \textbf{\# Sink\textsuperscript{1}} \\ 
\hline
Enterprise Collaboration (Enterprise) 
& General office agent scenarios for organizational collaboration. 
& WhatsApp, ClaudePost, Notes, Chrome 
& 43 & 17 & 6 \\ 
\hline
Software Dev \& IT Ops (DevOps)
& Professional agent scenarios for technical R\&D. 
& GitHub, DesktopCommander, Filesystem, MySQL 
& 67 & 18 & 12 \\ 
\hline
Social Platform Comm. (Social) 
& Social agent scenarios for individuals or small teams. 
& Chrome, RedNote, WhatsApp, Discord 
& 64 & 20 & 7 \\ 
\hline
Data Management \& Analysis (Data)
& Intelligent agent scenarios for data-driven tasks. 
& Filesystem, MySQL, Excel, QuickData 
& 59 & 18 & 4 \\ 
\hline
\textbf{Total} 
& - 
& - 
& \textbf{233} & \textbf{73} & \textbf{29} \\ 
\Xhline{1.1pt}
\end{tabularx}
\begin{flushleft}
\footnotesize
\textit{Note: \textsuperscript{1}A source refers to a function that retrieves user data and returns it to the Agent, whereas a sink refers to a function through which the Agent transmits data to a third party.}
\end{flushleft}
\end{table*}

%% file: Material/Tables/overall-result.tex
\begin{table}[htbp]
\caption{Overall effectiveness of \system{}.}
\label{table:overall-results}
\centering
\resizebox{\linewidth}{!}{
\renewcommand{\arraystretch}{1.25}
\begin{tabular}{l|ccccc}
\Xhline{1.1pt}
\textbf{Metric} 
& \textbf{Enterprise}
& \textbf{DevOps}
& \textbf{Social}
& \textbf{Data}
& \textbf{Total} \\
\hline
\textbf{Precision}
& \makecell{76 / 81 \\ (93.83\%)}
& \makecell{94 / 104 \\ (90.38\%)}
& \makecell{86 / 92 \\ (93.48\%)}
& \makecell{39 / 42 \\ (92.86\%)}
& \makecell{\textbf{295 / 319} \\ \textbf{(92.48\%)}} \\
\Xhline{1.1pt}
\end{tabular}
}
\flushleft{\footnotesize \textit{
Note: Precision is computed at the call-chain level as manually confirmed DOE-affected call chains over all call chains reported by \system{}.
}}
\end{table}

%% file: Material/Tables/CG-Construction.tex
\begin{table}[htbp]
\caption{Evaluation of Function Call Graph Construction.}
\label{table:call-graph-results}
\centering
\resizebox{\linewidth}{!}{
\renewcommand{\arraystretch}{1.3}
\large
\begin{tabular}{l|cc|c|c|c|c|c|c}
\Xhline{1.1pt}
\multirow{2}{*}{\textbf{Scenario}} 
& \multicolumn{2}{c|}{\textbf{Ground Truth FCG Size}} 
& \multirow{2}{*}{\textbf{TP}} 
& \multirow{2}{*}{\textbf{FP}} 
& \multirow{2}{*}{\textbf{FN}} 
& \multirow{2}{*}{\textbf{Precision (\%)}} 
& \multirow{2}{*}{\textbf{Recall (\%)}} 
& \multirow{2}{*}{\textbf{F1 (\%)}} \\ 
\cline{2-3}
& \textbf{Nodes} & \textbf{Edges} & & & & & & \\ 
\hline
Enterprise
& 43 & 144 & 136 & 2 & 8
& 98.55 & 94.44 & 96.45 \\
\hline
DevOps
& 67 & 216 & 188 & 11 & 28
& 94.47 & 87.04 & 90.60 \\
\hline
Social
& 64 & 207 & 186 & 6 & 21
& 96.88 & 89.86 & 93.23 \\
\hline
Data
& 59 & 153 & 143 & 23 & 10
& 86.14 & 93.46 & 89.66 \\
\hline
\textbf{Total}
& \textbf{233} & \textbf{720}
& \textbf{653} & \textbf{42} & \textbf{67}
& \textbf{93.96} & \textbf{90.69} & \textbf{92.30} \\
\Xhline{1.1pt}
\end{tabular}
}
\flushleft{\footnotesize \textit{Note: Metrics are computed at the functional-edge level. \textbf{TP}: correctly identified edges; \textbf{FP}: over-reported edges; \textbf{FN}: missed edges. TN is omitted as it represents the vast majority of non-existent edges in sparse call graphs and does not meaningfully reflect precision.}}
\end{table}

%% file: Material/Tables/User-Prompt-Generation.tex
\begin{table}[htbp]
\caption{User prompt synthesis and triggering effectiveness.}
\centering
\resizebox{\linewidth}{!}{
\renewcommand{\arraystretch}{1.3}
\large

\begin{tabular}{l|c|c|c}
\Xhline{1.1pt}
\textbf{Scenario} & \textbf{\# Generated Prompts} & \textbf{\# Valid Prompts} & \textbf{Trigger Coverage (\%)} \\ 
\hline
Enterprise    & 490  & 451  & 92.04 \\ 
\hline
DevOps      & 880 & 807 & 91.70 \\ 
\hline
Social       & 640  & 594  & 92.81 \\ 
\hline
Data & 360  & 355  & 98.61 \\ 
\hline
\textbf{Total}              & \textbf{2,370} & \textbf{2,207} & \textbf{93.12} \\ 
\Xhline{1.1pt}
\end{tabular}
}
\label{table:prompt-generation-high-perf}
\\
\end{table}

%% file: Material/Tables/DOE-Identification.tex
\begin{table}[htbp]
\caption{Evaluation of Over-Exposure Judge at the data-field level.}
\centering
\resizebox{\linewidth}{!}{
\renewcommand{\arraystretch}{1.3}
\large
\begin{tabular}{l|c|c|c|c|c|c|c}
\Xhline{1.1pt}
\textbf{Scenario}
& \textbf{TP} & \textbf{FP} & \textbf{FN} & \textbf{TN} & \textbf{Precision (\%)} & \textbf{Recall (\%)} & \textbf{F1 (\%)} \\ 
\hline
Enterprise    & 165 & 4 & 3 & 91 & 97.63 & 98.21 & 97.92 \\ 
\hline
DevOps    & 143 & 5 & 3 & 70 & 96.62 & 97.95 & 97.28 \\ 
\hline
Social       & 112 & 3 & 2 & 64 & 97.39 & 98.25 & 97.82 \\ 
\hline
Data & 198 & 4 & 3 & 91 & 98.02 & 98.51 & 98.26 \\ 
\hline
\textbf{Total}              & \textbf{618} & \textbf{16} & \textbf{11} & \textbf{316} & \textbf{97.48} & \textbf{98.25} & \textbf{97.86} \\ 
\Xhline{1.1pt}
\end{tabular}

}
\label{table:doe-judge-sampled}
\\
\flushleft{\footnotesize \textit{
\textbf{TP}: redundant fields correctly identified as DOE;
\textbf{FP}: necessary fields incorrectly flagged as DOE;
\textbf{TN}: necessary fields correctly identified as non-DOE;
\textbf{FN}: redundant fields missed by the judge.
Ground-truth labels are established through manual cross-verification of the transmitted fields.}}
\end{table}

%% file: Material/Tables/DOE-Identification-with-different-LLM.tex
\begin{table}[htbp]
\caption{Data-field-level comparison of single-LLM judges and multi-LLM voting in the Enterprise scenario (50 DOE-triggering and 50 non-DOE prompts).}
\label{table:doe-judge-comparison}
\centering
\resizebox{\linewidth}{!}{
\renewcommand{\arraystretch}{1.4}
\large
\begin{tabular}{l|c|c|c|c|c|c|c}
\Xhline{1.1pt}
\textbf{Method} & \textbf{TP} & \textbf{FP} & \textbf{FN} & \textbf{TN} & \textbf{Precision (\%)} & \textbf{Recall (\%) } & \textbf{F1 (\%)} \\ 
\hline
Single-LLM (GPT-4.1)          & 146 & 8 & 22 & 87 & 94.81 & 86.90 & 90.68 \\ 
\hline
Single-LLM (DeepSeek-V3.2)    & 149 & 13 & 19 & 82 & 91.98 & 88.69 & 90.30 \\ 
\hline
Single-LLM (Qwen3-Plus)       & 152 & 22 & 16 & 73 & 87.36 & 90.48 & 88.89 \\ 
\hline
\textbf{Ours (Multi-LLM Voting)} & \textbf{165} & \textbf{4} & \textbf{3} & \textbf{91} & \textbf{97.63} & \textbf{98.21} & \textbf{97.92} \\ 
\Xhline{1.1pt}
\end{tabular}
}
\\
\end{table}

%% file: Material/Tables/Impact-of-different-LLM-CG-and-UP.tex
\begin{table}[htbp]
\caption{Effectiveness of each individual module in different LLM models (Enterprise scenario).}
\centering
\resizebox{\linewidth}{!}{
\renewcommand{\arraystretch}{1.3}
\large

\begin{tabular}{l|cc|cc}
\Xhline{1.1pt}
\multirow{3}{*}{\textbf{LLM}} & \multicolumn{2}{c|}{\textbf{Call Graph Construction}} & \multicolumn{2}{c}{\textbf{User Prompt Synthesis}} \\ 
\cline{2-5} 
 & \textbf{F1 (\%)} & \begin{tabular}[c]{@{}c@{}}\textbf{\# Call Chains} \\ \textbf{involving DOE}\end{tabular} & \textbf{Coverage (\%)} & \begin{tabular}[c]{@{}c@{}}\textbf{\# User Prompts} \\ \textbf{involving DOE}\end{tabular} \\ 
\hline
GPT-4.1 & 85.40 & 66 & 87.14 & 253 \\ 
\hline
Qwen3-Plus$^\ddagger$ & 88.10 & 69 & \textbf{92.04} & \textbf{287} \\ 
\hline
DeepSeek-V3.2$^\dagger$ & \textbf{96.45} & \textbf{76} & 79.59 & 214 \\ 
\Xhline{1.1pt}
\end{tabular}
}
\label{table:model_selection}
\\
\flushleft{\footnotesize \textit{Note: $^\dagger$Selected for Call Graph Construction; $^\ddagger$Selected for User Prompt Generation. Bold values represent the peak performance that justifies our modular model selection strategy.}}
\end{table}

%% file: Material/Tables/overhead.tex
\begin{table}[t]
\caption{Overhead of \system{}, including resource efficiency, execution time, and coverage convergence (Scenario: Enterprise, Call chains involving DOE = 76).}
\label{table:cost-efficiency-final}
\centering
\footnotesize
\setlength{\tabcolsep}{1.15pt}
\renewcommand{\arraystretch}{1.12}
\resizebox{\columnwidth}{!}{%
\begin{tabular}{>{\raggedright\arraybackslash}p{0.29\columnwidth}|c|cc|c|c|c}
\Xhline{1.1pt}
\multirow{2}{*}{\textbf{Method}} & \multirow{2}{*}{\textbf{Coverage}} & \multicolumn{2}{c|}{\textbf{Execution Time (s)}} & \multirow{2}{*}{\textbf{Token Cost}} & \multirow{2}{*}{\textbf{Cost / Chain}} & \multirow{2}{*}{\shortstack{\textbf{Cost with} \\ \textbf{Full Coverage}}} \\ 
\cline{3-4}
 & & \textbf{Gen. Time} & \textbf{Det. Time} & & & \\ 
\hline
Variant 1 (w/o FCG, w/o Voting) & \raisebox{-0.6\height}{\shortstack{9.21\% \\ (7/76)}} & 118.14 & 1,488.43 & 0.43M & 0.061M & Infeasible$^*$ \\ 
\hline
Variant 2 (w/o FCG, w/ Voting)  & \raisebox{-0.6\height}{\shortstack{11.84\% \\ (9/76)}} & 118.14 & 1,534.77 & 1.16M & 0.129M & Infeasible$^*$ \\ 
\hline
Variant 3 (w/ FCG, w/o Voting)  & \raisebox{-0.6\height}{\shortstack{57.89\% \\ (44/76)}} & 128.93 & 1,497.06 & 0.42M & 0.010M & Total 8.2M \\ 
\hline
\textbf{Ours (w/ FCG, w/ Voting)} & \raisebox{-0.6\height}{\shortstack{\textbf{98.68\%} \\ \textbf{(75/76)}}} & \textbf{128.93} & \textbf{1,635.23} & \textbf{1.15M} & \textbf{0.015M} & \textbf{Total 1.2M} \\ 
\Xhline{1.1pt}
\end{tabular}
}
\\
\flushleft{\footnotesize \textit{Note: Coverage is measured over 150 generated prompts. Full-coverage cost is the total tokens required to identify all 76 DOE chains. $^*$Infeasible: no convergence within a 5,000-prompt budget owing to the lack of logic guidance. Gen./Det. Time denote prompt generation and environment execution plus DOE detection, respectively.}}
\end{table}

%% file: Material/Tables/realworld-results.tex
\begin{table*}[t]
\centering
\caption{DOE prevalence across real-world agents under the same backbone model and tool environment.}
\label{table:realworld-results}
\scriptsize
\setlength{\tabcolsep}{2.0pt}
\renewcommand{\arraystretch}{1.08}
\setlength{\aboverulesep}{0.15ex}
\setlength{\belowrulesep}{0.15ex}
\newcommand{\ixrowpad}{1.5pt}
\newcommand{\ixpaddedcell}[1]{%
  \raisebox{0pt}[\dimexpr\height+\ixrowpad\relax][\dimexpr\depth+\ixrowpad\relax]{\makecell{#1}}}
\newcommand{\res}[2]{\ixpaddedcell{#1\\[-0.1ex]{\footnotesize (#2)}}}
\newcommand{\best}[2]{\ixpaddedcell{\textbf{#1}\\[-0.1ex]{\footnotesize\textbf{(#2)}}}}

\resizebox{\textwidth}{!}{%
\begin{tabular}{l|ccc|ccc|ccc|ccc}
\toprule
\multirow{2}{*}{\textbf{Agent}}
& \multicolumn{3}{c|}{\textbf{Enterprise Collaboration}}
& \multicolumn{3}{c|}{\textbf{Software Dev \& IT Ops }}
& \multicolumn{3}{c|}{\textbf{Social Platform Comm.}}
& \multicolumn{3}{c}{\textbf{Data Management \& Analysis}} \\
\cmidrule(lr){2-4}
\cmidrule(lr){5-7}
\cmidrule(lr){8-10}
\cmidrule(l){11-13}
& \textbf{\makecell{Call\\Chain}}
& \textbf{\makecell{User\\Prompt}}
& \textbf{\makecell{Data\\Fields}}
& \textbf{\makecell{Call\\Chain}}
& \textbf{\makecell{User\\Prompt}}
& \textbf{\makecell{Data\\Fields}}
& \textbf{\makecell{Call\\Chain}}
& \textbf{\makecell{User\\Prompt}}
& \textbf{\makecell{Data\\Fields}}
& \textbf{\makecell{Call\\Chain}}
& \textbf{\makecell{User\\Prompt}}
& \textbf{\makecell{Data\\Fields}} \\
\midrule
\rowcolor[gray]{0.965}
OpenClaw
& \res{31/96}{32.29\%}
& \res{96/449}{21.38\%}
& \res{448/536}{83.58\%}
& \best{56/171}{32.75\%}
& \res{172/799}{21.53\%}
& \res{856/1,076}{79.55\%}
& \best{80/128}{62.50\%}
& \best{392/602}{65.12\%}
& \res{1,176/1,516}{77.57\%}
& \res{5/72}{6.94\%}
& \res{13/348}{3.74\%}
& \res{120/141}{85.11\%} \\
Hermes
& \res{52/97}{53.61\%}
& \res{180/453}{39.74\%}
& \res{623/808}{77.10\%}
& \res{44/169}{26.04\%}
& \res{131/797}{16.44\%}
& \res{631/855}{73.80\%}
& \res{34/127}{26.77\%}
& \res{168/594}{28.28\%}
& \res{503/646}{77.86\%}
& \best{13/71}{18.31\%}
& \best{36/354}{10.17\%}
& \res{366/416}{87.98\%} \\
\rowcolor[gray]{0.965}
Claude Code
& \res{22/94}{23.40\%}
& \res{70/447}{15.66\%}
& \res{128/220}{58.18\%}
& \res{8/175}{4.57\%}
& \res{18/809}{2.22\%}
& \res{88/128}{68.75\%}
& \res{52/127}{40.94\%}
& \res{164/590}{27.80\%}
& \res{460/641}{71.76\%}
& \res{3/72}{4.17\%}
& \res{5/356}{1.40\%}
& \res{35/43}{81.40\%} \\
Codex
& \res{45/97}{46.39\%}
& \res{219/457}{47.92\%}
& \res{875/1,098}{79.69\%}
& \res{54/174}{31.03\%}
& \best{241/807}{29.86\%}
& \best{1,493/1,840}{81.14\%}
& \res{68/128}{53.13\%}
& \res{275/592}{46.45\%}
& \best{1,236/1,508}{81.96\%}
& \res{5/72}{6.94\%}
& \res{8/355}{2.25\%}
& \best{220/245}{89.80\%} \\
\rowcolor[gray]{0.965}
Gemini CLI
& \best{62/96}{64.58\%}
& \best{252/456}{55.26\%}
& \best{855/1,010}{84.65\%}
& \res{51/172}{29.65\%}
& \res{193/802}{24.06\%}
& \res{1,053/1,330}{79.17\%}
& \res{65/127}{51.18\%}
& \res{242/589}{41.09\%}
& \res{858/1,073}{79.96\%}
& \res{9/72}{12.50\%}
& \res{21/353}{5.95\%}
& \res{195/226}{86.28\%} \\
\midrule
\rowcolor[gray]{0.90}
\textbf{Overall}
& \best{212/480}{44.17\%}
& \best{817/2,262}{36.12\%}
& \best{2,929/3,672}{79.77\%}
& \best{213/861}{24.74\%}
& \best{755/4,014}{18.81\%}
& \best{4,121/5,229}{78.81\%}
& \best{299/637}{46.94\%}
& \best{1,241/2,967}{41.83\%}
& \best{4,233/5,384}{78.62\%}
& \best{35/359}{9.75\%}
& \best{83/1,766}{4.70\%}
& \best{936/1,071}{87.39\%} \\
\bottomrule
\end{tabular}}

\begin{flushleft}
\scriptsize
\textit{Note: Each agent is evaluated with five synthesized prompts per executable taint path. ``Call Chain'' reports DOE-affected source-to-sink paths among those successfully triggered in at least one execution}; ``User Prompt'' reports DOE-triggering prompts among executions that successfully trigger the target call chain; ``Data Fields'' reports over-exposed fields among fields transmitted to sinks in DOE-triggering executions.
\end{flushleft}
\end{table*}